\documentclass[aip,apl,amsmath,amssymb,reprint]{revtex4-1}

\usepackage[utf8]{inputenc}
\usepackage[T1]{fontenc}
\usepackage{graphicx}
\usepackage{dcolumn}
\usepackage{bm}
\usepackage{color}
\usepackage{mathptmx}

\newcommand{\half}{\frac{1}{2}}

\newcolumntype{d}[1]{D{.}{.}{#1}}

\begin{document}

\preprint{AIP/123-QED}

\title[]{Uncertainty prediction in composite quantum chemistry approaches}

\author{Jakub Lang}
\author{Kacper Raczko}
\author{Micha\l\ Lesiuk}
\email{m.lesiuk@uw.edu.pl}
\affiliation{\sl University of Warsaw, Faculty of Chemistry, Pasteura 1, 02-093 Warsaw, Poland}

\date{\today}

\begin{abstract}
In this work, we consider the problem of uncertainty estimation of the results obtained through theoretical calculations using a composite quantum chemistry scheme. Each component of the composite scheme carries its own individual uncertainty, originating principally from a finite size of the basis set used in the calculations. Of practical interest is the total uncertainty, i.e. the uncertainty of the sum of all components. We first show that the conventional error propagation rules that treat each component as a statistically uncorrelated variable are not well-justified in practice. To remedy this, we propose a method of estimating the total uncertainty of the composite scheme which does not rely on the assumption that the components are independent. It is formulated as a series of random walks with different starting values that represent the uncertainty of each component of the composite scheme. This method is first applied to two example composite schemes for the water dimer and the nitrogen molecule that illustrate its most salient features and allow for a deeper analysis. Next, the method is tested for a larger set of interaction energies from the S66 dataset for which trustworthy reference data are available and the error can be assessed unambiguously. It is shown that the method provides reliable and reasonably tight uncertainty estimates at an arbitrary predefined confidence level required in a given application. While the focus of the paper is on composite schemes, we believe that the general idea can be useful more broadly in computational chemistry and physics.
\end{abstract}

\maketitle

\section{\label{sec:intro} Introduction}

Uncertainty quantification is an important task in many areas of science because any quantity obtained through measurement or through calculation carries an inherent error, absent some special circumstances. In comparing two or more estimates of the same quantity obtained from distinct sources and judging whether they are in agreement, one has to take into account their respective uncertainties. In experimental studies, the random error is usually assessed by repeating the same measurement numerous times and analyzing the results statistically, allowing one to define a confidence interval for the sample average. In addition, the results must be corrected for systematic errors in a given experimental setup and other factors~\cite{joint08}. 

In theoretical calculations, such an approach is usually impossible and one has to assess the uncertainty by other means~\cite{frombgen26}. The sources of errors in theoretical results can be categorized into several groups. According to the ``taxonomy'' proposed by Lejaeghere~\cite{lejaeghere20}, these are representation uncertainties, level of theory uncertainties, and numerics uncertainties. The first group takes into account the mismatch between what is measured and what is calculated, e.g. properties of a molecule in a vacuum instead of a gas phase, etc. The second group is related to the computational method used in the calculation, starting with the assumption of the Born-Oppenheimer approximation, neglect of relativistic effects, incomplete model of the electronic wavefunction, finite basis set size, etc. Finally, numeric uncertainties originate from the finite precision of the arithmetic used in the calculations, convergence thresholds, series truncation, and other factors of similar nature.

In this work, we focus on a situation in which the first and the third source of the uncertainty are practically negligible. In other words, we first assume that there is a near-perfect correspondence between the calculated and measured quantity. Second, we assume that the floating-point precision is sufficiently high to make numerical errors negligible, and there are no, e.g. convergence problems, that could spoil the accuracy of the calculated data. This leaves the level of theory as the only source of the error in this situation. Moreover, we focus on problems in which the Born-Oppenheimer approximation within a non-relativistic Schr\"odinger Hamiltonian is an entirely satisfactory computational model. While this choice is based mostly on the widespread use of this model, we see no significant obstacles in applying the main ideas of this work in the context of, e.g. the Dirac equation, or effective Hamiltonian models.

In principle, the non-relativistic Schr\"odinger equation for a molecule within the Born-Oppenherimer approximation can be solved using the full configuration interaction (FCI) method~\cite{siegbahn84,knowles84}. If this were possible, the only source of uncertainty would be the quality of the basis set used for the expansion of the molecular orbitals. If one knows FCI results in progression of systematically extended basis sets, the uncertainty of the data with respect to the complete basis set (CBS) limit can be assessed through the random walk uncertainty quantification procedure introduced in Ref.~\onlinecite{lang25} which would rigorously assign error bars at a given confidence level. Unfortunately, FCI calculations are feasible only for small molecules. To approximately reach the FCI limit in terms of the level of theory, one can apply a composite scheme~\cite{karton2016,peterson2012,feller2008,helgaker2008,martin2005}. In this approach, the quantity in question is calculated as a sum of corrections usually obtained as a combination of perturbation theory, coupled cluster methods~\cite{bartlett07,crawford07}, truncated configuration interaction, and/or density-functional theory. Note that these methods individually typically do not provide estimates of their uncertainty, with the notable exception of the density functionals based on a Bayesian approach~\cite{wellendorff2014,proppe2016,aldegunde2016}. However, the advantage of the composite idea is that the total quantity of interest is divided into a series of corrections decreasing in magnitude that sum up to near-FCI limit. The dominant corrections can be usually calculated using a lower-cost method within a larger basis set, whereas the more expensive terms are necessarily smaller and the use of a more economical basis set may be sufficient to represent them accurately.

The use of composite schemes (also known as protocols or recipes) in quantum chemistry is the most common in the context of thermochemistry or kinetics. Composite schemes such as Gaussian-$n$~\cite{pople1989,curtiss1990,curtiss1991,curtiss1998,curtiss2007}, HEAT~\cite{tajti2004,bomble2006,harding2008,thorpe2019}, Feller-Peterson-Dixon (FPD)~\cite{feller2008,feller2012,peterson2012,feller2013}, Weizmann-$n$~\cite{martin1999,boese2004,karton2006}, and correlation consistent Composite Approach~\cite{deyonker2006} (ccCA) are frequently used for high-accuracy predictions of atomization energies, reaction energies, etc., but it is impossible to cite all relevant contributions here. Composite schemes are also frequently used in accurate first-principles calculations of atomic and molecular properties useful in the fields of metrology~\cite{lesiuk20b,hellmann22,lang23,lang2023collision,lesiuk23,garberoglio23,przybytek26,lang26,lang26b}, ultracold chemistry and physics~\cite{tomza19,gronowski20,smialkowski21,ladjimi23,finelli24} or spectroscopy~\cite{lesiuk19,lesiuk20,gebala23,landau23,isolde25}. In these applications, the focus is typically on a single system or a small group of systems rather than on a broader set, and a proper composite scheme is usually defined on a case-by-case basis. However, independently of the application, in all composite schemes we face a similar challenge: having a well-defined set of contributions to the quantity of interest, we can estimate their respective uncertainties originating from finite basis set size using the aforementioned random walk procedure~\cite{lang25} or a different scheme. Then the question becomes how to assign a reliable uncertainty estimate to the sum of all these contributions. A commonly used approach to uncertainty prediction in composite schemes is to carry out calculations for a large data set of systems and compare the results with reliable external reference data obtained either from more accurate calculations or from experiments. However, such an approach is inherently not system specific, i.e. there is no guarantee that the conclusions hold for a system outside the initial dataset. Alternatively, one can assume that the components of the composite scheme are uncorrelated variables in the statistical sense, which allows one to determine the uncertainty of the sum solely from isolated uncertainties of the individual components using simple algebraic error propagation rules. However, in this work we argue that such an approach is not justified in general. Therefore, the goal of this work is to devise a system-specific and rigorous procedure to assign statistically meaningful uncertainty to a result obtained through a composite scheme without any reference to external data.

\section{\label{sec:random} Random walk uncertainty estimation}

\subsection{Outline of the procedure}

In this work, we build upon the uncertainty estimation procedure introduced in Ref.~\onlinecite{lang25}. For completeness, we outline the core of this method and its properties that are important from the point of view of this work.

Consider a certain quantity calculated in a progression of basis sets with the cardinal numbers $X=2,3,4,\ldots$ The value of this quantity obtained within the basis set $X$ is denoted by the symbol $E_X$. Knowing the results from two neighboring values of $X$, one can perform two-point extrapolation to the complete basis set (CBS) limit. We focus on the extrapolation scheme of Helgaker~\emph{et al.}~\cite{helgaker97,halkier98} as the most frequently used in the literature for this purpose, but other two-point extrapolation methods~\cite{martin96,lesiuk19b,varandas07,varandas18,varandas21} can also be used here.
Let us denote the extrapolated value obtained from the pair of basis sets with cardinal numbers ($X$,$X-1$) by $e_X$. The fundamental assumption of the method from Ref.~\onlinecite{lang25} is that the differences between neighboring values of $e_X$ are decreasing as a function of $X$. In other words, for each $X$ we have the lower and upper bounds:
\begin{align}
\label{eq:fundamental-bound}
    e_X - |e_X - e_{X-1}| < e_{X+1} < e_X + |e_X - e_{X-1}|.
\end{align}
Note that to fulfill this constraint, the individual values of $e_X$ (or $E_X$) do not necessarily have to follow any consistent pattern. Even if $e_X$ or $E_X$ exhibit an oscillatory or irregular convergence pattern, the differences $e_{X+1} - e_X$ can still conform to the bound~(\ref{eq:fundamental-bound}). As shown in Ref.~\onlinecite{lang25} the exceptions from this bound are rare in practice, and modifications of the procedure that deal with this problem were discussed therein.

Assume that we have calculated two extrapolated values, $e_X$ and $e_{X-1}$, and determination of the corresponding results for larger $X$ is unfeasible due to prohibitive computational costs or other factors. We do not know what the next extrapolated value $e_{X+1}$ is but we know that it has to obey the bound~(\ref{eq:fundamental-bound}). Therefore, we simply randomize the next value $\tilde{e}_{X+1}$ from the interval~(\ref{eq:fundamental-bound}). This procedure is then iterated and the next $\tilde{e}_{X+2}$ is randomized from the interval:
\begin{align}
    \tilde{e}_{X+1} - |\tilde{e}_{X+1} - e_X| < \tilde{e}_{X+2} < \tilde{e}_{X+1} + |\tilde{e}_{X+1} - e_X|.
\end{align}
This procedure converges fast and after a finite number of steps two consecutive values $\tilde{e}_{X+N}$ and $\tilde{e}_{X+N+1}$ differ in absolute terms by less than a predefined numerical threshold. This gives us a numerically converged value $\tilde{e}_\infty$, which is an estimate of the CBS limit of the quantity in question. Note that this procedure is essentially a random walk in which the length of the next step depends on the outcome of the two previous steps.

Of course, a single realization of the described procedure is physically meaningless. It represents only one
possible scenario of what could have happened if results in larger basis sets had been available. However, when a large number of such simulations are performed with the same starting conditions $e_X$ and $e_{X-1}$, obtaining a different value of $\tilde{e}_\infty$ each time, the distribution of $\tilde{e}_\infty$ can be analyzed statistically. This naturally leads to system-specific uncertainty estimates for $\tilde{e}_\infty$ expressed through confidence intervals, as demonstrated in Ref.~\onlinecite{lang25}.

Note that the outlined procedure requires results from three consecutive basis sets, $E_X$, $E_{X-1}$, and $E_{X-2}$, to generate the initial extrapolated results, $e_X$ and $e_{X-1}$. However, in a situation where results from only two basis sets are available, most commonly $E_2$ and $E_3$, we propose to employ the pair $e_3$ and $E_3$ to kickstart the random walk procedure. Finally, when a result from only one basis set is available, it is impossible to use the random walk scheme, and one has to resort to other methods to assign an uncertainty to such data.

\subsection{The distribution of $\tilde{e}_\infty$}

Note that we do not know mathematically at this point what the distribution of $\tilde{e}_\infty$ obtained in the limit of an infinite number of independent simulations is. This was not a significant problem from the point of view of Ref.~\onlinecite{lang25} because the focus was always on obtaining the CBS limit of a single quantity $E$. Therefore, it was sufficient to carry out a large number of independent simulations with the same starting values $e_X$ and $e_{X-1}$, generate the histogram of the results, and calculate all the necessary statistical parameters of the distribution from this histogram. With around a million simulations, the obtained numerical approximation of the distribution was sufficiently accurate for all practical purposes, and the time required for this task was on the order of a few seconds. However, in this work, we are interested in providing uncertainty estimates for composite quantum chemistry schemes. In the next section, we shall show that this problem can be solved if we perform the described random walk procedure for a large number of different starting values $e_X$ and $e_{X-1}$. Assuming that we need a set of million starting values to get stable results and each combination of starting values requires a million realizations of the procedure, the computational costs would no longer be manageable. Therefore, in this section, we analyze the mathematical properties of the distribution of $\tilde{e}_\infty$ and calculate its relevant properties analytically. This will drastically simplify the simulations for each fixed combination of starting values $e_X$ and $e_{X-1}$.

To simplify the notation, denote the random variable $\tilde{e}_\infty$ by the symbol $E$. We are interested in the exact probability distribution $\rho_E$ of this random variable obtained in the limit of an infinite number of realizations of the random walk procedure. From Ref.~\onlinecite{lang25} we only know that this distribution is symmetric around its mean, which is equal to $e_X$. Let us introduce a new random variable $Y$ that is related to $E$ through the formula:
\begin{align}
\label{eq:shifty}
    E = e_X + Y ( e_X - e_{X-1} ),
\end{align}
where $e_X$ and $e_{X-1}$ are the two values used to initialize the random walk. Denote the probability distribution of the random variable $Y$ by the symbol $\rho_Y$. As a result of the shifting and scaling in Eq.~(\ref{eq:shifty}), the distribution $\rho_Y$ is universal in the sense that the initial values for the random walk of $Y$ are $y_X=0$ and $y_{X-1}=-1$, both independent of $e_X$ and $e_{X-1}$. This is a considerable simplification that apparently has not been realized in Ref.~\onlinecite{lang25}.

The universal distribution $\rho_Y$ is much easier to analyze analytically. In Appendix~\ref{sec:appendix} we show that it obeys the following distributional equation:
\begin{align}
\label{eq:master}
    Y = U(1+Y),
\end{align}
and this leads to the integral representation:
\begin{align}
\label{eq:rhoint}
    \rho_Y(y) = \frac{1}{\pi} \int_{0}^{+\infty} dt\;e^{-\mbox{\scriptsize Cin}(t)}\,\cos(ty),
\end{align}
where $\mbox{Cin}(t)$ is the cosine integral function. This equation enables us to probe the probability distribution $\rho_Y$ without the Monte-Carlo procedure and hence bypass the painfully slow convergence with respect to the number of samples. To evaluate Eq.~(\ref{eq:rhoint}) numerically, we adopted the \verb|quadosc| function as implemented in the \verb|mpmath| Python library for handling oscillatory integrals\cite{mpmath}. The calculations were carried out in extended arithmetic precision and the number of decimal places used in the calculations was increased until convergence of the results to at least eight significant digits was observed. We generated the probability distribution in steps of $10^{-5}$ within the interval $(0,6.2)$. Outside this interval the value of $\rho_Y$ drops below $10^{-8}$ and hence it is practically negligible.

The numerical integration is useful within the whole domain of $\rho_Y(y)$ apart from the region of small $y$, where it requires unreasonably high arithmetic precision. Therefore, for $|y|\leq\frac{1}{100}$ we use the following mixed power-logarithmic series expansion:
\begin{align}
\label{eq:rhoyexp}
    \rho_Y(y) = b_0 + b_2 y^2 + a_0 \ln|y| + a_2 y^2 \ln|y| + \ldots
\end{align}
where the values of the coefficients $a_n$ and $b_n$ are given in Appendix~\ref{sec:appendix}. The error of this formula is $\mathcal{O}(y^4\ln|y|)$ in the leading order and hence negligible for such small $y$.

\section{\label{sec:independence} Statistical dependence in composite schemes}

With the help of the random walk procedure~\cite{lang25} described in the previous section, we are able to assign statistically meaningful error bars to a single quantity obtained through theoretical calculations in a progression of basis sets, followed by an extrapolation to the CBS limit. However, in composite quantum chemistry methods, the quantity of interest, such as the interaction energy, atomization energy, etc., is calculated as the sum of contributions obtained at different levels of theory. Each contribution can be extrapolated to the CBS limit, and the random walk procedure would assign error bars at a predefined confidence level. The main question then becomes how to obtain reliable and statistically meaningful error bars for the whole sum.

The most common approach in the literature to this problem is to assume that each contribution is a statistically independent random variable. This drastically simplifies the problem because for two independent variables $X$ and $Y$, the probability distribution of their sum $\rho_{X+Y}$ is a convolution of the probability distributions for the two individual components, i.e. $\rho_X*\rho_Y$. As a result, the variance, confidence intervals, and other important parameters characterizing the probability distribution of the sum can be directly calculated from their counterparts for the two components. For example, if two components $X$ and $Y$ are normally distributed with variances $\sigma_X^2$ and $\sigma_Y^2$ then the variance for the sum is simply $\sigma_{X+Y}^2 = \sigma_X^2 + \sigma_Y^2$. For other probability distributions, such relationships would be more complicated, especially for confidence intervals, but ultimately the knowledge of $\rho_X$ and $\rho_Y$ is entirely sufficient to find all useful properties of the random variable $X+Y$. This is no longer true if the variables $X$ and $Y$ are statistically dependent. In this case, one has to consider, e.g. the covariance between $X$ and $Y$ in evaluation of various metrics of the probability of the sum as stressed by Pernot and Savin~\cite{pernot18,pernot20}.

Therefore, it is first instructive to study whether the assumption that the contributions in the composite schemes are independent holds in practice. To this end, we consider the interaction energies of the molecular complexes from the S66 dataset~\cite{rezac11}, popular in benchmarking various quantum chemistry methods. The number of molecules included in the set (66) is large enough to draw statistically meaningful conclusions. We consider the composite scheme named ``Silver'' designed by Kesharwani and collaborators~\cite{kesharwani18}. It consists of the following four contributions to the total interaction energy of each system:
\begin{enumerate}
    \item Hartree-Fock contribution calculated within aug-cc-pVQZ-F12 basis set~\cite{peterson08} with addition of a perturbative correction for the basis set incompleteness~\cite{adler07,noga09};
    \item MP2 correlation contribution calculated using the MP2-F12 method~\cite{werner07} within aug-cc-pVTZ-F12 and aug-cc-pVQZ-F12 basis set pair~\cite{sylvetsky17}, and extrapolated to the CBS limit;
    \item difference between CCSD \cite{purvis82} and MP2 contributions, calculated using CCSD(F12*) \cite{tew07} and MP2-F12 methods within aug-cc-pVTZ-F12 basis set;
    \item contribution of perturbative triples (T) correction~\cite{raghavachari89} calculated with haug-cc-pVDZ and haug-cc-pVTZ basis sets \cite{dunning89,kendall92} and extrapolated to the CBS limit.
\end{enumerate}
We follow the recommendation from Ref.~\onlinecite{kesharwani18} and rely on the half-counterpoise corrected \cite{boys70} data. The raw data used for the present analysis, as well as further technical details of the calculations, are found in the Supplementary Material of Ref.~\onlinecite{kesharwani18}. Note that in the main text of Ref.~\onlinecite{kesharwani18} the first two contributions are summed together, but here we split them into separate components due to their physically distinct nature. For brevity, we abbreviate the four contributions defined above as $E_{\mathrm{HF}}$, $E_{\mathrm{MP2}}$, $E_{\mathrm{SD}}$, and $E_{\mathrm{(T)}}$ in order of appearance.

There are many metrics by which we could measure the dependence between two random variables. The simplest one is the Pearson correlation coefficient, denoted $r$, which measures the linear dependence between two datasets. The values of the Pearson correlation coefficient $r\approx 1$ or $r \approx -1$ signify a strong direct proportionality with positive or negative coefficient, respectively, which is the simplest form of dependence. Therefore, by calculating the Pearson correlation coefficient between the $E_{\mathrm{HF}}$, $E_{\mathrm{MP2}}$, $E_{\mathrm{SD}}$, and $E_{\mathrm{(T)}}$ contributions for the S66 dataset, we can judge the linear dependence between them. If we find a statistically significant linear dependence, the hypothesis that the components of the composite scheme introduced above are independent is decisively negated. However, note that the converse statement is not necessarily true. Even if the Pearson correlation coefficient is exactly zero, this does not exclude dependence in a non-linear form. Nevertheless, for size-extensive quantities such as the interaction energies, non-linear correlations are less likely.

\begin{table}
\caption{\label{tab:pearson}
Pearson correlation coefficients $r$ between components of the ``Silver'' composite scheme from Ref.~\onlinecite{kesharwani18}. 
}
\begin{ruledtabular}
\begin{tabular}{c|cccccc}
  & $E_{\mathrm{HF}}$ & $E_{\mathrm{MP2}}$ & $E_{\mathrm{SD}}$ & $E_{\mathrm{(T)}}$ &
    $E_{\mathrm{HF+MP2}}$ & $E_{\mathrm{SD(T)}}$ \\
\hline
$E_{\mathrm{HF}}$      & \phantom{$-$}1.000 & \\
$E_{\mathrm{MP2}}$     & $-$0.498 & \phantom{$-$}1.000 \\
$E_{\mathrm{SD}}$      & \phantom{$-$}0.524 & $-$0.937 & \phantom{$-$}1.000 \\
$E_{\mathrm{(T)}}$     & $-$0.261           & \phantom{$-$}0.960 & $-$0.854 & \phantom{$-$}1.000 \\
$E_{\mathrm{HF+MP2}}$  & \phantom{$-$}0.854 & \phantom{$-$}0.025 & \phantom{$-$}0.042 &
 \phantom{$-$}0.274 & \phantom{$-$}1.000 \\
$E_{\mathrm{SD(T)}}$   & \phantom{$-$}0.608 & $-$0.801 & \phantom{$-$}0.954 & $-$0.658 & 
\phantom{$-$}0.220 & \phantom{$-$}1.000 \\
\end{tabular}
\end{ruledtabular}
\end{table}

In Table~\ref{tab:pearson} we show the values of the Pearson correlation coefficients calculated between the components $E_{\mathrm{HF}}$, $E_{\mathrm{MP2}}$, $E_{\mathrm{SD}}$, and $E_{\mathrm{(T)}}$ of the ``Silver'' composite scheme from Ref.~\onlinecite{kesharwani18}. For future discussion, we also added to the table sums of the first two components ($E_{\mathrm{HF}}+E_{\mathrm{MP2}}$ denoted $E_{\mathrm{HF+MP2}}$) and the remaining two components ($E_{\mathrm{SD}}+E_{\mathrm{(T)}}$ denoted $E_{\mathrm{SD(T)}}$) as separate entries. Consider the correlation between the four basic components $E_{\mathrm{HF}}$, $E_{\mathrm{MP2}}$, $E_{\mathrm{SD}}$, and $E_{\mathrm{(T)}}$. First, the correlation between Hartree-Fock and all remaining contributions is rather weak ($|r|<0.6$). This is not entirely surprising, because the Hartree-Fock and correlation contributions are based on distinct physical descriptions of electron-electron interactions (mean field vs. instantaneous Coulomb field). However, all correlation contributions are strongly correlated with each other. In particular, $E_{\mathrm{MP2}}$ and $E_{\mathrm{SD}}$ are strongly inversely correlated ($r\approx-0.937$), and similarly for $E_{\mathrm{SD}}$ and $E_{\mathrm{(T)}}$ ($r\approx-0.854$). Simultaneously, the correlation coefficient $r\approx 0.960$ between $E_{\mathrm{MP2}}$ and $E_{\mathrm{(T)}}$ is a strong indicator of proportionality. In contrary, there are some components or combinations thereof for which no correlation is found. For example, the pair $E_{\mathrm{HF+MP2}}$ and $E_{\mathrm{SD}}$ shows essentially no correlation ($r\approx 0.42$) and similarly for $E_{\mathrm{HF+MP2}}$ and $E_{\mathrm{(T)}}$ ($r\approx 0.274$).

Overall, we find that the assumption that components of the composite schemes are independent (in the statistical sense) is not justified, as exemplified by the example of the S66 dataset. Therefore, simple rules of error propagation are generally not safe to use when evaluating the total uncertainty for a quantity calculated with the composite scheme. As a result, we have two ways to proceed. First, we can attempt to design composite schemes in which the components are grouped in a particular way so that the correlation between them is small. For example, for the ``Silver'' composite scheme from Ref.~\onlinecite{kesharwani18} we could group the components into $E_{\mathrm{HF+MP2}}$ and $E_{\mathrm{SD(T)}}$, because the correlation between them is essentially zero ($r\approx 0.220$). However, it is not clear whether the conclusions found here for the S66 benchmark set translate into a more general phenomenon that holds equally well for other properties and molecular systems. Moreover, such a combination of different quantities may not always be optimal from the computational standpoint. For example, for large molecules the $E_{\mathrm{SD}}$ component can usually be calculated within a larger basis set than $E_{\mathrm{(T)}}$ due to the steeper scaling of the latter method ($N^7$ vs. $N^6$). By combining them together into $E_{\mathrm{SD(T)}}$ we are forced to ``downgrade'' to a smaller basis set and this largely defeats the purpose of a composite scheme. Alternatively, we can abandon the assumption that the components in a composite scheme are truly independent in the statistical sense and seek uncertainty estimation methods that bypass this requirement. This is the route taken in this work.

\section{\label{sec:uqcomposite} Uncertainty estimation in the composite schemes}

\subsection{A simple example}

\begin{table}
\caption{\label{tab:waterdim}
Correlation contributions $E_{\mathrm{MP2}}$ and $E_{\mathrm{SD}}$ to the interaction energy of the water dimer (in kJ/mol) calculated within the aug-cc-pV$X$Z basis set family. The extrapolation to the CBS limit is performed using the two-point formula of Helgaker~\emph{et al.} The confidence intervals reported in the last row correspond to the 95\% certainty level in the random walk procedure.
}
\begin{ruledtabular}
\begin{tabular}{ccc}
 $X$ & $E_{\mathrm{MP2}}$ & $E_{\mathrm{SD}}$  \\
\hline
\multicolumn{3}{c}{Raw contributions} \\
\hline
2 & 3.224 & $-$0.873 \\
3 & 4.599 & $-$0.782 \\
4 & 5.133 & $-$0.795 \\
5 & 5.330 & -- \\
\hline
\multicolumn{3}{c}{$(X,X-1)$ extrapolated results} \\
\hline
3 & 5.178 & $-$0.744 \\
4 & 5.523 & $-$0.805 \\
5 & 5.536 & -- \\
\hline
\multicolumn{3}{c}{Best CBS estimates} \\
\hline
-- & 5.536 $\pm$ 0.020 & $-$0.805 $\pm$ 0.092 \\
\end{tabular}
\end{ruledtabular}
\end{table}

To introduce the proposed uncertainty estimation procedure for the composite schemes, we begin with a simple example that illustrates the core logic of the method. In the next section we will generalize this to a more complicated example and finally discuss the general framework.

Assume that we wish to calculate the best theoretical estimate of the interaction energy of the water dimer at the CCSD level of theory. The structure of the water dimer used in this example is given in the Supplementary Material \cite{supp} (in the XYZ format). We introduce a simple two-component composite scheme that consists of (i) the MP2 correlation contribution to the interaction energy, denoted $E_{\mathrm{MP2}}$, and (ii) the difference between CCSD and MP2 correlation contributions, denoted $E_{\mathrm{SD}}$. The practical reason for this partitioning is the lower computational cost of the MP2 method compared to CCSD, such that the $E_{\mathrm{MP2}}$ component can typically be calculated within a larger basis set than $E_{\mathrm{SD}}$. Note that we consider pure correlation energy contributions at both MP2 and CCSD levels, because the Hartree-Fock contribution to the interaction energy is relatively straightforward to calculate and converges much faster with the basis set size than the dynamical correlation effects. Therefore, for simplicity, we assume that the HF contribution is known and has effectively negligible uncertainty.

In Table~\ref{tab:waterdim} we report the values of the contributions $E_{\mathrm{MP2}}$ and $E_{\mathrm{SD}}$ to the interaction energy of the water dimer calculated with aug-cc-pV$X$Z basis set family with $X=2,3,4,5$. We introduce a shorthand notation $E_X^{\mathrm{MP2}}$ and $E_X^{\mathrm{SD}}$ to refer to the values of these contributions calculated for a given $X$. We adopt the sign convention that the contribution to the interaction energy is positive or negative when it is attractive or repulsive, respectively. For the sake of demonstration, we assume that MP2 results are available up to $X=5$ and CCSD results up to $X=4$, and calculations with larger basis sets are unfeasible for both components. All results were corrected for the basis set superposition error (BSSE) using the conventional counterpoise scheme. For each pair of results from two consecutive basis sets $(X,X-1)$ we performed a two-point extrapolation to the CBS limit using the scheme of Helgaker~\emph{et al.}~\cite{helgaker97,halkier98} The results are also included in Table~\ref{tab:waterdim}. Further in the text, we denote the extrapolated values with the symbols $e_X^{\mathrm{MP2}}$ and $e_X^{\mathrm{SD}}$. Finally, we employ the random walk procedure for each of the two components $E_{\mathrm{MP2}}$ and $E_{\mathrm{SD}}$ separately to determine the corresponding confidence intervals at the 95\% certainty level. Other levels of certainty could also be considered here, but we focus on the 95\% in this demonstration as the choice sufficient in most practical cases.

At this point, we know the best theoretical estimates for the components $E_{\mathrm{MP2}}$ and $E_{\mathrm{SD}}$ in the complete basis set limit and their respective uncertainties. Our goal is to find an uncertainty for the sum of these components at the same confidence level. From the discussion given in Sec.~\ref{sec:independence} we know that we cannot simply assume that $E_{\mathrm{MP2}}$ and $E_{\mathrm{SD}}$ are independent statistical variables. Therefore, conventional error propagation rules, such as summing the squares of the individual uncertainties and taking the square root, are not well justified. Without any additional considerations, the only legitimate option would simply be to take the sum of the two individual uncertainties. However, this corresponds to the worst-case scenario when both components are highly positively correlated and the errors of both components add up. This is statistically unlikely, especially that we observe a partial cancellation between $E_{\mathrm{MP2}}$ and $E_{\mathrm{SD}}$, and hence some degree of cancellation between their errors is also expected. In such a situation, adding both uncertainties is a safe option, but would likely lead to a gross overestimation of the actual uncertainty at a fixed confidence level.

The main difficulty in estimating the uncertainty of the sum of $E_{\mathrm{MP2}}$ and $E_{\mathrm{SD}}$ originates from the fact that for the latter the extrapolated result from the basis set pair $(4,5)$, i.e. $e_5^{\mathrm{SD}}$, is not available. If this additional information were available, there would be no issue: we could add the contributions $E_X^{\mathrm{MP2}}$ and $E_X^{\mathrm{SD}}$ together \emph{before} performing extrapolations and then estimate the uncertainty using the random walk procedure only once, i.e. directly the sum of the two quantities. However, since $e_5^{\mathrm{SD}}$ is missing, we propose to apply the same fundamental idea as in the random walk procedure, i.e. to fall back to the bounds expressed in Eq.~(\ref{eq:fundamental-bound}). Under the same assumptions, the value of $e_5^{\mathrm{SD}}$ is within the interval:
\begin{align}
\label{eq:e5bound}
    e_4^{\mathrm{SD}} - | e_4^{\mathrm{SD}} - e_3^{\mathrm{SD}} | < e_5^{\mathrm{SD}} < e_4^{\mathrm{SD}} + | e_4^{\mathrm{SD}} - e_3^{\mathrm{SD}} |,
\end{align}
and can be simply drawn from a uniform distribution, giving $\tilde{e}_5^{\mathrm{SD}}$. Once this is done, one can add the entire set of extrapolated contributions $e_X^{\mathrm{MP2}}$ and $e_X^{\mathrm{SD}}$ together (with $e_5^{\mathrm{SD}}$ replaced by a randomized $\tilde{e}_5^{\mathrm{SD}}$). This generates two starting values for the random walk procedure: $e_5^{\mathrm{(tot)}}=e_5^{\mathrm{MP2}}+\tilde{e}_5^{\mathrm{SD}}$ and $e_4^{\mathrm{(tot)}}=e_4^{\mathrm{MP2}}+e_4^{\mathrm{SD}}$. With these fixed values, we run the random walk $N$ times and generate the distribution of the CBS limit for the sum. Of course, results obtained with a single set of randomized starting values are meaningless. However, when we perform the entire procedure $M$ times, i.e. each time generating new randomized starting values, we amass $M\times N$ estimates of the CBS limit for the sum, $e_\infty^{\mathrm{(tot)}}$. The data are then analyzed statistically to find the confidence interval for $e_\infty^{\mathrm{(tot)}}$, i.e. an interval that covers a predefined percentage of the data points around the mean of the distribution. However, as we shall see shortly, the distribution obtained as the outcome of this procedure is not necessarily symmetric around the mean. As a result, the upper and lower bounds of the confidence interval are not necessarily equidistant from the average value. Therefore, to find the central confidence interval at a given $p\%$ confidence level, we have to find two intervals: the first covering $\half p\%$ of the data points with values lower than the mean, and separately the second covering $\half p\%$ of the data points above the mean. 

\begin{figure}
    \centering\hspace{-0.5cm}
    \includegraphics[width=1.0\linewidth]{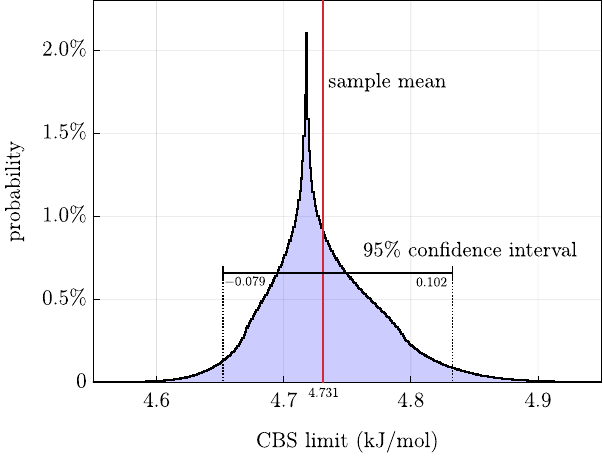}
    \caption{Histogram illustrating results of the simulations for the interaction energy of water dimer (in kJ/mol) with $M=10^6$ random initializations and $N=10^6$ random walks for each initial value. The red vertical line represents the mean result obtained from $M\cdot N$ samples. The vertical brackets represent the confidence interval at 95\% confidence level.}
    \label{fig:water-hist}
\end{figure}

In Fig.~\ref{fig:water-hist} we report the results of the simulations for the simple two-component composite scheme for the interaction energy of the water dimer based on the data from Table~\ref{tab:waterdim}. The results correspond to $M=10^6$ randomized initial values and $N=10^6$ independent random walks for each starting value. A further increase in this parameter leads to no appreciable changes in the uncertainty predictions and only tiny fluctuations of the distribution. In the histogram plotted in Fig.~\ref{fig:water-hist} we use a relatively small width of the bin ($10^{-3}$) which makes the shape of the graph almost continuous. Several important features of the distribution become apparent from the shape of the plot in Fig.~\ref{fig:water-hist}. First, the mean value of the distribution is almost exactly equal to the sum of the best CBS estimates of both components ($e_5^{\mathrm{MP2}} + e_4^{\mathrm{SD}}$) with the value $5.536 - 0.805 = 4.371\,$kJ/mol, see the last row of Table~\ref{tab:waterdim}. In fact, the mean value of the distribution differs from this sum at the seventh significant digit, and this tiny discrepancy can be attributed to the finite, albeit huge, sample size $M$ and $N$. Therefore, the proposed procedure enables us to estimate the uncertainty of the best CBS estimate of the quantity in question, but does not affect the value itself. The second feature of the distribution in Fig.~\ref{fig:water-hist} is the pronounced asymmetry. In fact, the right wing of the distribution has a significantly longer tail than the left wing. As a consequence, the maximum of the distribution is not equal to its mean value. Additionally, the endpoints of the confidence intervals obtained from analysis of the histogram are not equidistant from the mean. In this example, we focus on the 95\% certainty level which yields the confidence interval $[-0.079,0.102]$ (centered at the mean). Therefore, the best estimates of the CBS limit of the calculated quantity and its uncertainty should be reported as $4.371_{-0.079}^{+0.102}\,$kJ/mol at the 95\% confidence level. 

Let us compare the obtained interval with other methods for assigning uncertainty to the calculated data. As discussed above, the most pessimistic approach is to take the simple sum of the uncertainties of both components. Knowing $5.536\,\pm\,0.020$ and $-0.805\,\pm\, 0.092$ for the MP2 and CCSD correlation contributions from Table~\ref{tab:waterdim}, this gives $4.371\,\pm\,0.112$ for the sum. The error bars obtained in this way are necessarily symmetric around the best CBS estimate and are also significantly broader than the confidence interval obtained from the proposed procedure. This is not surprising as this method does not account for any possibility of error cancellation and assumes the worst case scenario of perfect correlation where the errors simply add up. The second method of assigning uncertainties assumes that the two components are statistically independent. In such a situation, it is justified to take a sum of squares of the individual uncertainties and take the square root. This leads to the estimate $4.371\,\pm\,0.094$, which also predicts a symmetric distribution of the uncertainty. Compared with the proposed method, the lower bound of the confidence interval is overestimated, while the upper bound is slightly underestimated.

The advantage of the simple example considered in this section is the fact that the main results can be confirmed and rationalized analytically. To this end, we find expressions for the starting values in each random walk procedure. First, recall that the quantity $\tilde{e}_5^{\mathrm{SD}}$ is randomized according to Eq.~(\ref{eq:e5bound}) and therefore can be written as $\tilde{e}_5^{\mathrm{SD}} = e_4^{\mathrm{SD}} + u | e_4^{\mathrm{SD}} - e_3^{\mathrm{SD}} |$, where $u$ is a random variable with uniform distribution within the interval $[-1,1]$. Therefore, each random walk is initialized with the starting value $e_5^{\mathrm{(sum)}} = e_5^{\mathrm{MP2}} + \tilde{e}_5^{\mathrm{SD}} = e_5^{\mathrm{MP2}} + e_4^{\mathrm{SD}} + u | e_4^{\mathrm{SD}} - e_3^{\mathrm{SD}} |$. For notational simplicity, we write this as $e_5^{\mathrm{(sum)}} = A + Bu$, where $A = e_5^{\mathrm{MP2}} + e_4^{\mathrm{SD}}$ and $B = | e_4^{\mathrm{SD}} - e_3^{\mathrm{SD}} |$. The second quantity required to initiate the random walk is the difference between the two consecutive extrapolated values for the sum, i.e. $e_5^{\mathrm{(sum)}} - e_4^{\mathrm{(sum)}} = e_5^{\mathrm{MP2}} - e_4^{\mathrm{MP2}} + \tilde{e}_5^{\mathrm{SD}} - e_4^{\mathrm{SD}} = e_5^{\mathrm{MP2}} - e_4^{\mathrm{MP2}} +  u | e_4^{\mathrm{SD}} - e_3^{\mathrm{SD}} |$. Further in the text, we write this as $e_5^{\mathrm{(sum)}} - e_4^{\mathrm{(sum)}} = C + Bu$, where $C=e_5^{\mathrm{MP2}} - e_4^{\mathrm{MP2}}$ and $B$ is the same as above. For a fixed value of $u$, the results obtained within a single random walk are described by the probability distribution $\rho_E(x)$ which reads
\begin{align}
\begin{split}
    \rho_E(x) &= \frac{1}{|e_5^{\mathrm{(sum)}} - e_4^{\mathrm{(sum)}}|}\,\rho_Y\left( \frac{x - e_5^{\mathrm{(sum)}}}{e_5^{\mathrm{(sum)}} - e_4^{\mathrm{(sum)}}} \right) = \\
    &= \frac{1}{|C + Bu|}\,\rho_Y\left( \frac{x - A - Bu}{C + Bu} \right)
\end{split}
\end{align}
according to Eq.~(\ref{eq:shifty}) and the accompanying discussion from Sec.~\ref{sec:random}. To find the distribution $\bar{\rho}_E(x)$ obtained from a collection of random walks with different values of $u$, directly corresponding to the histogram in Fig.~\ref{fig:water-hist}, we have to average $\rho_E(x)$ over all possible $u$, weighted by the probability distribution of $u$. As $u$ is uniformly distributed within $[-1,1]$, this leads to the integral representation:
\begin{align}
\label{eq:barrhoe}
\begin{split}
    &\bar{\rho}_E(x) = \half \int_{-1}^{+1} du\; \rho_E(x) = \\
    &= \half \int_{-1}^{+1} du\; \frac{1}{|C + Bu|}\,\rho_Y\left( \frac{x - A - Bu}{C + Bu} \right).
\end{split}
\end{align}
With this knowledge, we can find the mean value $\mu$ of the distribution $\bar{\rho}_E(x)$ which corresponds to the red vertical line in Fig.~\ref{fig:water-hist}. From the definition, the mean value of $\bar{\rho}_E(x)$ is given by the integral
\begin{align}
\begin{split}
    &\mu = \int_{-\infty}^{+\infty} dx\;x\,\bar{\rho}_E(x) = \\
    &\half \int_{-\infty}^{+\infty} dx\;x\int_{-1}^{+1} du\; 
    \frac{1}{|C + Bu|}\,\rho_Y\left( \frac{x - A - Bu}{C + Bu} \right).
\end{split}
\end{align}
To simplify this expression, we switch the order of integrations, change the integration variable from $x$ to $y=\frac{x - A - Bu}{|C + Bu|}$ and use the fact that the distribution $\rho_Y$ is symmetric around its mean (zero). This leads to:
\begin{align*}
\begin{split}
    &\mu = \half \int_{-1}^{+1} du\; \int_{-\infty}^{+\infty} dy\;( y|C + Bu| + A + Bu )
    \rho_Y(y) = \\
    &= \half \int_{-1}^{+1} du\; \Big[ |C + Bu|\int_{-\infty}^{+\infty} dy\; y\,\rho_Y(y) \\
    &+ (A + Bu )\int_{-\infty}^{+\infty} dy\;
    \rho_Y(y) \Big].
\end{split}
\end{align*}
The two inner integrals over $y$ are elementary. The first is equal to zero, $\int_{-\infty}^{+\infty} dy\; y\,\rho_Y(y)=0$, because the distribution $\rho_Y(y)$ has zero mean, while the second is equal to unity, $\int_{-\infty}^{+\infty} dy\; \rho_Y(y)=1$, from the normalization condition of $\rho_Y(y)$. The mean value simplifies to:
\begin{align}
    \mu = \half \int_{-1}^{+1} du\; ( A + Bu ),
\end{align}
and the integration over $u$ is elementary, giving $\mu = A = e_5^{\mathrm{MP2}} + e_4^{\mathrm{SD}}$. This confirms the numerical result from Fig.~\ref{fig:water-hist} and proves that the mean value of the distribution simply corresponds to the best available theoretical estimate for the sum of both components.

From Eq.~(\ref{eq:barrhoe}) one can also justify why the confidence intervals calculated from the numerical data are not distributed symmetrically around the mean. This can be traced back to the fact that $\bar{\rho}_E$ does not possess any symmetry, unlike $\rho_Y$. Indeed, from Eq.~(\ref{eq:barrhoe}) we see that $\bar{\rho}_E$ is symmetric around its mean ($A$) only when $B=0$ or $C=0$. Recalling the definition of these parameters, the condition $B=0$ would imply that $e_4^{\mathrm{SD}} = e_3^{\mathrm{SD}}$, while $C=0$ means that $e_5^{\mathrm{MP2}} = e_4^{\mathrm{MP2}}$. Neither of these equalities holds for the data in Table~\ref{tab:waterdim} and is also unlikely to be encountered in practice.

Finally, this analysis also enables us to justify why the confidence intervals are broader on the right-hand side of the mean than on the left-hand side, as seen from our final estimate obtained above, $4.371_{-0.079}^{+0.102}\,$kJ/mol. To see this, we recall that the starting values for each random walk are $e_5^{\mathrm{(sum)}} = e_5^{\mathrm{MP2}} + e_4^{\mathrm{SD}} + u | e_4^{\mathrm{SD}} - e_3^{\mathrm{SD}} |$ and $e_5^{\mathrm{(sum)}} - e_4^{\mathrm{(sum)}} = e_5^{\mathrm{MP2}} - e_4^{\mathrm{MP2}} +  u | e_4^{\mathrm{SD}} - e_3^{\mathrm{SD}} |$ for a given value of $u$. The former parameter, $e_5^{\mathrm{(sum)}}$, corresponds to the mean value of the distribution obtained from the collection of random walks with fixed starting values, while the latter, $e_5^{\mathrm{(sum)}} - e_4^{\mathrm{(sum)}}$, dictates how broad the obtained distribution is. To obtain a mean value larger than $A=e_5^{\mathrm{MP2}} + e_4^{\mathrm{SD}}$, the value of the parameter $u$ must be positive. In such a situation, $e_5^{\mathrm{(sum)}} - e_4^{\mathrm{(sum)}}$ is greater than both $e_5^{\mathrm{MP2}} - e_4^{\mathrm{MP2}}$ and $| e_4^{\mathrm{SD}} - e_3^{\mathrm{SD}} |$ individually, as seen from the definition above, i.e. there is no cancellation between these quantities. In contrast, to obtain a mean value smaller than $A=e_5^{\mathrm{MP2}} + e_4^{\mathrm{SD}}$, the value of $u$ must be negative. Under this condition, the terms $e_5^{\mathrm{(MP2)}} - e_4^{\mathrm{(MP2)}}$ and $| e_4^{\mathrm{SD}} - e_3^{\mathrm{SD}} |$ in the expression for $e_5^{\mathrm{(sum)}} - e_4^{\mathrm{(sum)}}$ partially cancel out, making the distribution on the left-hand side of the mean narrower than on the right-hand side. To summarize this conclusion in simple terms, the proposed algorithm automatically recognizes that to obtain the mean value smaller than the best theoretical estimate, i.e. $e_5^{\mathrm{MP2}} + e_4^{\mathrm{SD}}$, the exact values of the two components (which are not available directly, but are simulated for each $u$) have to partially cancel out compared to $e_5^{\mathrm{MP2}} + e_4^{\mathrm{SD}}$. Under this condition, it is statistically likely that the uncertainties of both components will also cancel out to some degree. On the other hand, if the mean value becomes greater than $e_5^{\mathrm{MP2}} + e_4^{\mathrm{SD}}$, the probability of such errors cancellation is less likely.

\subsection{An extended example}

\begin{table}
\caption{\label{tab:n2int}
Correlation contributions $E_{\mathrm{fc-SD}}$, $E_{\mathrm{fc-T}}$, $E_{\mathrm{fc-Q}}$, and $E_{\mathrm{ae-T}}$ to the interaction energy of N$_2$ molecule (in eV). The extrapolation to the CBS limit is performed using two-point formula of Helgaker~\emph{et al.}. The confidence intervals reported in the last row correspond to the 99\% certainty level in the random walk procedure.
}
\begin{ruledtabular}
\begin{tabular}{ccccc}
 $X$ & $E_{\mathrm{fc-SD}}$ & $E_{\mathrm{fc-T}}$ & $E_{\mathrm{fc-Q}}$ & $E_{\mathrm{ae-T}}$ \\
\hline
\multicolumn{3}{c}{Raw contributions} \\
\hline
2 & --     & --     & 0.0408 & -- \\
3 & --     & --     & 0.0411 & 0.0269 \\
4 & --     & 0.3611 & 0.0425 & 0.0332 \\
5 & --     & 0.3672 & --     & 0.0359 \\
6 & 4.1873 & 0.3692 & --     &  \\
7 & 4.2041 & --     & --     &  \\
8 & 4.2134 & --     & --     &  \\
\hline
\multicolumn{3}{c}{$(X,X-1)$ extrapolated results} \\
\hline
3 & --     & --     & 0.0413 & -- \\
4 & --     & --     & 0.0435 & 0.0378 \\
5 & --     & 0.3736 & --     & 0.0388 \\
6 & --     & 0.3719 & --     & -- \\
7 & 4.2327 & --     & --     & -- \\
8 & 4.2322 & --     & --     & -- \\
\hline
\multicolumn{5}{c}{Best CBS estimates} \\
\hline
-- & 4.2322$\,\pm\,$0.0010 & 0.3719$\,\pm\,$0.0035 & 0.0435$\,\pm\,$0.0045 & 0.0388$\,\pm\,$0.0020 \\
\end{tabular}
\end{ruledtabular}
\end{table}

Let us now move to a more complicated example where the proposed procedure can be applied. We consider the interaction energy in the N$_2$ molecule at the internuclear distance $R=2.070\,$a.u. (which is very close to the equilibrium bond length) based on the calculations reported in Ref.~\onlinecite{lang26}. In the composite scheme considered in this example, the total interaction energy is split into five components: 
\begin{itemize}
    \item Hartree-Fock contribution, $E_{\mathrm{HF}}$;
    \item frozen-core CCSD correlation contribution, $E_{\mathrm{fc-SD}}$;
    \item difference between frozen-core CCSDT and CCSD contributions, $E_{\mathrm{fc-T}}$;
    \item difference between frozen-core CCSDTQ and CCSDT contributions, $E_{\mathrm{fc-Q}}$;
    \item contribution of core electrons evaluated with the CCSDT method, $E_{\mathrm{ae-T}}$;
\end{itemize}
In the frozen-core calculations, the $1s$ orbitals of each nitrogen atom were not correlated. The last contribution, $E_{\mathrm{ae-T}}$, was evaluated as a difference in the interaction energy between frozen-core approach and with all-electrons correlated using the CCSDT method. All frozen-core contributions were evaluated using the aug-cc-pV$X$Z basis family, while the $E_{\mathrm{ae-T}}$ term was obtained within the aug-cc-pCV$X$Z counterparts. Similarly as before, we assume that the Hartree-Fock contribution carries a marginal error, and we neglect it from further considerations, leaving only the remaining four components that bring a significant uncertainty. The raw results for all four components are taken from Ref.~\onlinecite{lang26} and are included in Table~\ref{tab:n2int} for completeness. In this table, we also provide extrapolated data using two-point scheme of Helgaker~\emph{et al.}~\cite{helgaker97,halkier98}. As the composite scheme considered in this section was designed with a higher precision in mind compared to the simple example from the previous section, we increase the target certainty level to 99\%. The confidence intervals for each of the four components of the interaction energy determined using the random walk procedure at this confidence level are reported in Table~\ref{tab:n2int}. Our goal is to find the confidence interval for the sum of all four components at the same confidence level.

\begin{figure}
    \centering\hspace{-0.5cm}
    \includegraphics[width=1.0\linewidth]{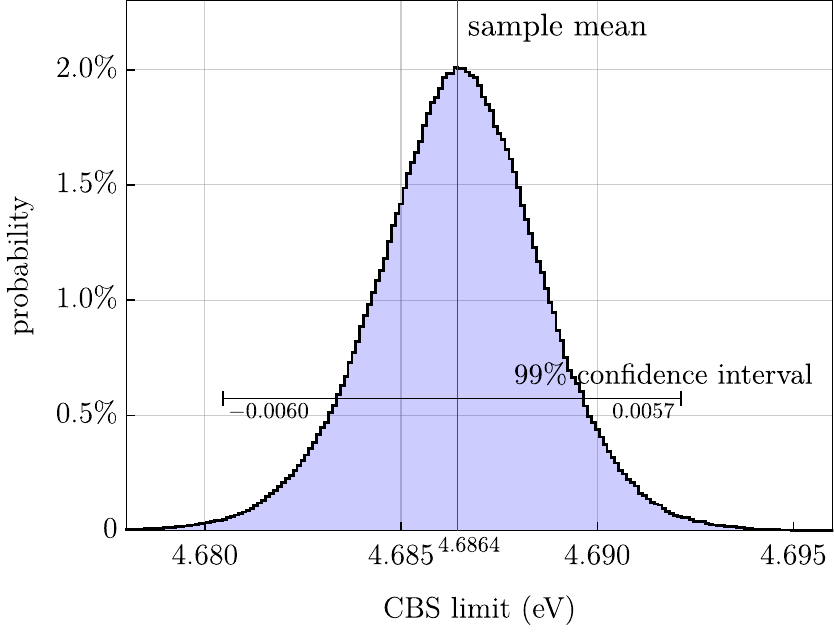}
    \caption{Histogram illustrating results of the simulations for the interaction energy of N$_2$ molecule (in eV) with $M=10^6$ random initializations and $N=10^6$ random walks for each initial value. The red vertical line represents the mean result obtained from $M\cdot N$ samples. The vertical brackets represent the confidence interval at 99\% confidence level.}
    \label{fig:n2-hist}
\end{figure}

The example we consider here is more complicated only because we have to consider four components that constitute the quantity of interest. Another problem stems from the fact that many more data points are missing as many results are not available in larger basis sets due to the prohibitive computational cost of some corrections. For example, while for $E_{\mathrm{fc-SD}}$ results up to $X=8$ are available, only $X=4$ is feasible for $E_{\mathrm{fc-Q}}$. To fill in the missing entries, we propose to iterate the randomization procedure outlined in the previous section. For example, in the case of $E_{\mathrm{fc-T}}$ we know $e_5$ and $e_6$, but $e_7$ and $e_8$ are not available. To find $e_7$ we invoke the fundamental assumption that differences between extrapolated values vanish monotonically and hence we have a bound:
\begin{align}
    e_6 - |e_6 - e_5| < e_7 < e_6 + |e_6 - e_5|.
\end{align}
Therefore, we can randomize $e_7$ from this interval and get $\tilde{e}_7$. With this knowledge, we repeat the same process for $e_8$: write the analogous bounds for $e_8$ using the $\tilde{e}_7$ generated previously and randomize $\tilde{e}_8$ from the resulting interval. In this case, two random draws are necessary to fill in the missing values. An analogous procedure can be performed for the remaining components. The main difference is that for $E_{\mathrm{fc-Q}}$ four randomization steps are necessary to complete the table up to $e_8$, while three are needed for $E_{\mathrm{ae-T}}$. Once the missing results are generated by repeating the randomization procedure within the appropriate bound, the entire set of results can be added together to generate a set of candidate values for the sum. This enables us to run the random walk procedure $N$ times and generate a distribution of possible results originating from the given starting values. This procedure is then repeated $M$ times, each time with different starting values, and from the $M\cdot N$ data points, a histogram of the results can be assembled. The rest of the procedure is identical as in the simple example considered in the previous section and as a final outcome, a confidence interval at the specified confidence level is determined.

In Fig.~\ref{fig:n2-hist} we present the results of the simulations for $M=N=10^6$ independent samples. A striking feature of the histogram presented in Fig.~\ref{fig:n2-hist} is that it is nearly symmetric around the mean. In fact, the deviation from strict symmetry can be assessed numerically but is barely perceptible at the scale of the plot, in sharp contrast to the data presented in the previous section. Moreover, the distribution carries a strong resemblance to the normal distribution. This translates into a more symmetric distribution of the confidence intervals. At the 99\% confidence level, the confidence interval is $[-0.0060,0.0057]$ and therefore the final result should be reported as $4.6864^{+0.0057}_{-0.0060}\,$eV. This can be compared with the aforementioned standard methods for assigning uncertainties. In the most pessimistic approach, we take a simple sum of the uncertainties of individual components, giving $4.6864\pm 0.0110\,$eV. The confidence interval obtained in this way is much broader. The difference is more striking than in the simple example considered in the previous section, which is a simple consequence of the fact that the number of components in the composite scheme is larger. We can also compare our result with the confidence interval obtained by summing squares of the uncertainties and taking the square root, i.e. assuming that the components are statistically independent. This leads to $4.6864\pm 0.0061\,$eV. In this case, the confidence interval is much closer to that obtained from the proposed method.

The ``extended example'' suggests that when the number of components in the composite scheme is large, the distribution eventually converges to a Gaussian one, even if these components are correlated. While this statement appears to be a manifestation of the central limit theorem, we have not managed to prove it mathematically and we doubt the generality of this observation. Moreover, even if this statement is true, under some assumptions that are fulfilled in practice, one frequently encounters composite schemes in which only two or three components bring a dominant contribution to the total error. As the previous ``simple example'' clearly shows, the distribution of the results can be very far from a Gaussian in such a case. 

\section{Further applications}

\subsection{Implementation details}

In the example applications reported in Sec.~\ref{sec:uqcomposite}, we used a brute-force approach where initial values for the random walk procedure are randomly generated $M$ times, and subsequently $N$ samples are used in each random walk loop. With both $M$ and $N$ being of an order of a million, necessary to generate numerically converged results, a huge total number of samples ($M\cdot N$) was required. This is why the calculations reported in Sec.~\ref{sec:uqcomposite} took about two weeks of continuous runtime, even after a parallelization of the sampling procedure was applied. Clearly, this is not a way forward, because the proposed uncertainty quantification scheme would be, in many cases, more computationally intensive than the underlying electronic structure calculations, adding a substantial overhead from the point of view of large-scale applications.

To reduce the cost of the proposed procedure (without deteriorating the results) we use the analytical results obtained for the probability distribution $\rho_Y$ in Sec.~\ref{sec:random}. Taking advantage of the fact that this distribution is universal, in the sense discussed above, we can drastically reduce the timings of the simulations as follows. First of all, in our reduced-cost implementation we generate $M$ composite tables of the randomized initial values of $e_{X}^{\mathrm{(tot)}}$ using the procedure described in Sec.~\ref{sec:uqcomposite}. By default, $M=10^6$ is used. This step is identical to the first step in the na\"{i}ve approach and, unfortunately, cannot be bypassed. In the na\"{i}ve approach, we would further proceed with $N$ random walk estimations of $e_{\infty}^{\mathrm{(tot)}}$ for each generated table i.e. $N\cdot M$ random walks in total. Instead of this, we calculate  $e_{\infty}^{\mathrm{(tot)}}$ directly for each table using the probability $\rho(y)$. The internal probability $\tilde{\rho}(y)$ is created by joining the analytical solution Eq.~(\ref{eq:rhoyexp}) for $|y|\leq\frac{1}{100}$ and the interpolation of the numerical solutions of Eq.~\ref{eq:rhoint} for $|y|>\frac{1}{100}$. This probability distribution is then binned into an internal histogram with bin width equal to one tenth of the user specified value of the global histogram. The default bin width for the global histogram is $10^{-4}$ and it governs the final resolution of the data. In our numerical experience, the default settings provide uncertainties stable to at least two significant digits, which is entirely sufficient in most cases. The population of each bin is then calculated as a definite integral of $\tilde{\rho}(y)$ between two edges of each bin. Using the estimated uncertainties, we rescale the bins of the internal histogram by $e_{X}^{\mathrm{(tot)}} - e_{X-1}^{\mathrm{(tot)}}$ and shift the center of the histogram by the corresponding value of $e_{X}^{\mathrm{(tot)}}$. These modified histograms are then binned into the global coarser histogram, collecting the global information across all samples. To speed up the process, the rescaling, shifting and rebinning are processed for a bulk of tables and are heavily vectorized with the only size restriction being due to the memory footprint. Finally, the global histogram is then used to estimate the confidence intervals and the sample mean.

By applying the simplifications of the proposed procedure described above, enabled by the knowledge of the universal distribution $\rho_Y$ discussed in Sec.~\ref{sec:random}, it is possible to reduce the computational cost of the uncertainty estimation procedure for a single system from about two weeks to around 7 hours (with one million randomized initial values). Such an overhead would likely be acceptable in most application of the composite schemes used in the literature. In the special case where only a single component of a composite scheme is present, the first step described in the previous paragraph is completely skipped and the global histogram is directly tied to a single internal histogram rescaled by $e_{X}-e_{X-1}$ and shifted by $e_{X}$. This yields results fully consistent with the method proposed in Ref.~\onlinecite{lang25}.

A Python implementation of the proposed uncertainty quantification procedure, along with a short description of its use, is available on GitHub \cite{github}. Example input files that can be used to reproduce the results reported in Sec.~\ref{sec:uqcomposite} are also provided. Additionally, the implementation contains the raw values of the probability distribution $\rho_Y$ that were determined according to the procedure described in Sec.~\ref{sec:random}.

\subsection{\label{subsec:composite_s66}Illustrative composite scheme}

\begin{table*}
\caption{\label{tab:s66_1}
Uncertainties of the results obtained with the composite scheme from Sec.~\ref{subsec:composite_s66} for the interaction energies for complexes from the S66 dataset. The ``14k-Gold'' scheme from Ref.~\onlinecite{nagy23} is used as the reference (reproduced in the last column) while the actual errors with respect to the reference are given in the second last column. The confidence intervals are reported at 75\%, 95\%, and 99\% confidence levels (the third, fourth, and fifth column, respectively). The most narrow confidence interval that is successful in uncertainty prediction is given in bold.
All results are given in kcal/mol.
}
\begin{ruledtabular}
\begin{tabular}{cd{2.4}cccrd{2.4}}
 System & \multicolumn{1}{c}{Result} & 
 \multicolumn{3}{c}{Confidence interval} & 
 Error & \multicolumn{1}{c}{Reference} \\
 \hline
        &        & 75\% & 95\% & 99\% & \\
 \hline
  1 &  5.006 & \textbf{[$-$0.025, 0.022]} & [$-$0.041, 0.046] & [$-$0.054, 0.067] & $-$0.015 &  4.991 \\
  2 &  5.701 & \textbf{[$-$0.030, 0.026]} & [$-$0.048, 0.055] & [$-$0.062, 0.080] & $-$0.027 &  5.674 \\
  3 &  7.028 & [$-$0.032, 0.030] & \textbf{[$-$0.055, 0.060]} & [$-$0.075, 0.086] & $-$0.032 &  6.996 \\
  4 &  8.212 & \textbf{[$-$0.028, 0.029]} & [$-$0.054, 0.052] & [$-$0.077, 0.071] & $-$0.021 &  8.191 \\
  5 &  5.860 & [$-$0.029, 0.026] & \textbf{[$-$0.045, 0.053]} & [$-$0.056, 0.078] & $-$0.033 &  5.827 \\
\hline
  6 &  7.661 & [$-$0.032, 0.029] & \textbf{[$-$0.052, 0.059]} & [$-$0.066, 0.087] & $-$0.035 &  7.625 \\
  7 &  8.344 & [$-$0.029, 0.026] & \textbf{[$-$0.048, 0.053]} & [$-$0.064, 0.078] & $-$0.031 &  8.313 \\
  8 &  5.088 & \textbf{[$-$0.021, 0.019]} & [$-$0.034, 0.039] & [$-$0.043, 0.057] & $-$0.017 &  5.071 \\
  9 &  3.115 & [$-$0.023, 0.020] & \textbf{[$-$0.036, 0.042]} & [$-$0.046, 0.062] & $-$0.023 &  3.091 \\
 10 &  4.213 & [$-$0.023, 0.026] & \textbf{[$-$0.047, 0.043]} & [$-$0.069, 0.056] & $-$0.024 &  4.189 \\
\hline
 11 &  5.467 & [$-$0.018, 0.020] & \textbf{[$-$0.036, 0.032]} & [$-$0.054, 0.039] & $-$0.025 &  5.442 \\
 12 &  7.388 & \textbf{[$-$0.032, 0.033]} & [$-$0.061, 0.061] & [$-$0.086, 0.085] & $-$0.028 &  7.360 \\
 13 &  6.279 & [$-$0.024, 0.024] & \textbf{[$-$0.045, 0.045]} & [$-$0.063, 0.062] & $-$0.029 &  6.250 \\
 14 &  7.550 & [$-$0.029, 0.028] & \textbf{[$-$0.052, 0.053]} & [$-$0.072, 0.075] & $-$0.033 &  7.518 \\
 15 &  8.715 & [$-$0.022, 0.024] & \textbf{[$-$0.044, 0.042]} & [$-$0.064, 0.057] & $-$0.027 &  8.688 \\
\hline
 16 &  5.193 & \textbf{[$-$0.017, 0.017]} & [$-$0.032, 0.032] & [$-$0.044, 0.045] & $-$0.015 &  5.178 \\
 17 & 17.420 & \textbf{[$-$0.052, 0.056]} & [$-$0.105, 0.098] & [$-$0.151, 0.133] & $-$0.017 & 17.403 \\
 18 &  6.956 & [$-$0.024, 0.027] & \textbf{[$-$0.050, 0.045]} & [$-$0.074, 0.058] & $-$0.027 &  6.929 \\
 19 &  7.497 & [$-$0.021, 0.023] & \textbf{[$-$0.043, 0.039]} & [$-$0.062, 0.052] & $-$0.032 &  7.465 \\
 20 & 19.427 & \textbf{[$-$0.064, 0.061]} & [$-$0.114, 0.120] & [$-$0.156, 0.171] & $-$0.044 & 19.383 \\
\hline
 21 & 16.531 & \textbf{[$-$0.062, 0.056]} & [$-$0.105, 0.114] & [$-$0.141, 0.165] & $-$0.046 & 16.485 \\
 22 & 19.778 & \textbf{[$-$0.057, 0.057]} & [$-$0.107, 0.107] & [$-$0.150, 0.149] & $-$0.031 & 19.748 \\
 23 & 19.459 & \textbf{[$-$0.060, 0.058]} & [$-$0.109, 0.112] & [$-$0.151, 0.159] & $-$0.035 & 19.424 \\
 24 &  2.674 & \textbf{[$-$0.050, 0.044]} & [$-$0.082, 0.091] & [$-$0.107, 0.134] &  0.013 &  2.686 \\
 25 &  3.744 & \textbf{[$-$0.049, 0.044]} & [$-$0.081, 0.090] & [$-$0.106, 0.131] &  0.015 &  3.760 \\
\hline
 26 &  9.629 & \textbf{[$-$0.105, 0.112]} & [$-$0.208, 0.179] & [$-$0.310, 0.231] &  0.087 &  9.716 \\
 27 &  3.289 & \textbf{[$-$0.050, 0.045]} & [$-$0.082, 0.092] & [$-$0.107, 0.136] &  0.015 &  3.304 \\
 28 &  5.486 & \textbf{[$-$0.073, 0.078]} & [$-$0.146, 0.138] & [$-$0.210, 0.187] &  0.060 &  5.545 \\
 29 &  6.597 & \textbf{[$-$0.072, 0.079]} & [$-$0.148, 0.134] & [$-$0.216, 0.179] &  0.064 &  6.661 \\
 30 &  1.340 & \textbf{[$-$0.024, 0.024]} & [$-$0.044, 0.045] & [$-$0.062, 0.063] &  0.003 &  1.343 \\
\hline
 31 &  3.281 & \textbf{[$-$0.034, 0.037]} & [$-$0.068, 0.058] & [$-$0.100, 0.073] &  0.025 &  3.306 \\
 32 &  3.644 & \textbf{[$-$0.032, 0.036]} & [$-$0.067, 0.059] & [$-$0.098, 0.076] &  0.031 &  3.675 \\
 33 &  1.776 & \textbf{[$-$0.025, 0.024]} & [$-$0.045, 0.046] & [$-$0.062, 0.066] &  0.009 &  1.784 \\
\end{tabular}
\end{ruledtabular}
\end{table*}

\begin{table*}
\caption{\label{tab:s66_2}
The same as Table~\ref{tab:s66_1} but for the remaining systems from the S66 dataset.
}
\begin{ruledtabular}
\begin{tabular}{cd{2.4}cccrd{2.4}}
 System & \multicolumn{1}{c}{Result} & 
 \multicolumn{3}{c}{Confidence interval} & 
 Error & \multicolumn{1}{c}{Reference} \\
 \hline
        &        & 75\% & 95\% & 99\% & \\
 \hline
 34 &  3.755 & \textbf{[$$-$$0.022, 0.025]} & [$-$0.046, 0.039] & [$-$0.068, 0.048] & $-$0.022 &  3.733 \\
 36 &  1.773 & [$-$0.009, 0.010] & [$-$0.019, 0.018] & \textbf{[$-$0.028, 0.024]} & $-$0.022 &  1.750 \\
 37 &  2.391 & [$-$0.013, 0.013] & \textbf{[$-$0.025, 0.025]} & [$-$0.035, 0.035] & $-$0.016 &  2.375 \\
 38 &  2.964 & \textbf{[$-$0.017, 0.015]} & [$-$0.027, 0.032] & [$-$0.033, 0.047] & $-$0.008 &  2.957 \\
\hline
 39 &  3.475 & \textbf{[$-$0.019, 0.017]} & [$-$0.031, 0.035] & [$-$0.042, 0.051] &  0.007 &  3.482 \\
 40 &  2.831 & \textbf{[$-$0.014, 0.014]} & [$-$0.027, 0.027] & [$-$0.037, 0.039] & $-$0.007 &  2.823 \\
 41 &  4.749 & \textbf{[$-$0.045, 0.045]} & [$-$0.086, 0.080] & [$-$0.124, 0.111] &  0.020 &  4.768 \\
 42 &  4.028 & \textbf{[$-$0.035, 0.036]} & [$-$0.068, 0.061] & [$-$0.101, 0.083] &  0.026 &  4.055 \\
 43 &  3.648 & \textbf{[$-$0.034, 0.035]} & [$-$0.066, 0.062] & [$-$0.096, 0.086] &  0.012 &  3.660 \\
\hline
 44 &  1.983 & \textbf{[$-$0.015, 0.015]} & [$-$0.028, 0.026] & [$-$0.040, 0.036] & $-$0.007 &  1.976 \\
 45 &  1.706 & \textbf{[$-$0.008, 0.009]} & [$-$0.017, 0.014] & [$-$0.025, 0.017] & $-$0.003 &  1.704 \\
 46 &  4.226 & \textbf{[$-$0.019, 0.019]} & [$-$0.037, 0.036] & [$-$0.051, 0.050] & $-$0.010 &  4.216 \\
 47 &  2.788 & \textbf{[$-$0.023, 0.021]} & [$-$0.040, 0.043] & [$-$0.053, 0.063] &  0.018 &  2.806 \\
 48 &  3.466 & \textbf{[$-$0.020, 0.018]} & [$-$0.032, 0.037] & [$-$0.041, 0.054] &  0.011 &  3.477 \\
\hline
 49 &  3.248 & \textbf{[$-$0.022, 0.021]} & [$-$0.040, 0.041] & [$-$0.055, 0.058] &  0.018 &  3.267 \\
 50 &  2.835 & \textbf{[$-$0.025, 0.024]} & [$-$0.041, 0.047] & [$-$0.057, 0.068] &  0.001 &  2.836 \\
 51 &  1.528 & \textbf{[$-$0.003, 0.003]} & [$-$0.005, 0.005] & [$-$0.006, 0.008] &  0.000 &  1.528 \\
 52 &  4.707 & [$-$0.013, 0.012] & \textbf{[$-$0.022, 0.025]} & [$-$0.030, 0.036] & $-$0.016 &  4.691 \\
 53 &  4.385 & [$-$0.004, 0.004] & [$-$0.007, 0.008] & [$-$0.008, 0.011] & $-$0.010 &  4.375 \\
\hline
 54 &  3.277 & [$-$0.009, 0.009] & \textbf{[$-$0.016, 0.016]} & [$-$0.023, 0.023] & $-$0.014 &  3.263 \\
 55 &  4.156 & [$-$0.016, 0.016] & \textbf{[$-$0.027, 0.031]} & [$-$0.037, 0.045] & $-$0.019 &  4.137 \\
 56 &  3.178 & \textbf{[$-$0.011, 0.010]} & [$-$0.018, 0.021] & [$-$0.024, 0.030] & $-$0.007 &  3.171 \\
 57 &  5.218 & \textbf{[$-$0.019, 0.017]} & [$-$0.032, 0.035] & [$-$0.043, 0.050] &  0.000 &  5.219 \\
 58 &  4.203 & \textbf{[$-$0.028, 0.030]} & [$-$0.055, 0.047] & [$-$0.083, 0.059] & $-$0.009 &  4.193 \\
\hline
 59 &  2.916 & \textbf{[$-$0.019, 0.021]} & [$-$0.038, 0.033] & [$-$0.057, 0.041] & $-$0.003 &  2.913 \\
 60 &  4.955 & \textbf{[$-$0.023, 0.023]} & [$-$0.039, 0.044] & [$-$0.053, 0.065] & $-$0.022 &  4.933 \\
 61 &  2.890 & \textbf{[$-$0.017, 0.017]} & [$-$0.031, 0.032] & [$-$0.044, 0.044] & $-$0.011 &  2.879 \\
 62 &  3.508 & \textbf{[$-$0.019, 0.020]} & [$-$0.037, 0.037] & [$-$0.052, 0.051] & $-$0.013 &  3.495 \\
 63 &  3.712 & \textbf{[$-$0.030, 0.032]} & [$-$0.059, 0.056] & [$-$0.085, 0.077] &  0.004 &  3.716 \\
\hline
 64 &  2.984 & \textbf{[$-$0.010, 0.010]} & [$-$0.018, 0.018] & [$-$0.025, 0.025] & $-$0.007 &  2.977 \\
 65 &  4.083 & \textbf{[$-$0.029, 0.032]} & [$-$0.058, 0.050] & [$-$0.087, 0.061] & $-$0.012 &  4.071 \\
 66 &  3.950 & [$-$0.005, 0.006] & [$-$0.010, 0.011] & [$-$0.014, 0.016] & $-$0.020 &  3.931 \\
\end{tabular}
\end{ruledtabular}
\end{table*}

In this section, we will apply the proposed scheme for uncertainty estimation to a larger set of systems. The goal is to judge whether the proposed scheme indeed provides reliable confidence intervals at the specified uncertainty level by comparing it with accurate reference data.

To this end, we consider the CBS limit of the CCSD(T) interaction energies for complexes from the S66 database~\cite{rezac11}. The reference data for this quantity is taken from Ref.~\onlinecite{nagy23} (``14k-Gold'' scheme) and is assumed to have an uncertainty of less than 0.01 kcal/mol, which is acceptable for our purposes. To approximately reproduce the reference results, we designed an economical composite scheme with the following components:
\begin{enumerate}
    \item Hartree-Fock contribution calculated within aug-cc-pVQZ-F12 basis set with addition of a perturbative correction for the basis set incompleteness;
    \item MP2 correlation contribution calculated within haug-cc-pVQZ and haug-cc-pV5Z basis set pair, and extrapolated to the CBS limit;
    \item difference between CCSD(T) and MP2 contributions calculated within haug-cc-pVDZ and haug-cc-pVTZ basis sets, and extrapolated to the CBS limit.
\end{enumerate}
All extrapolations were performed using the two-point $X^{-3}$ scheme of Helgaker~\emph{et al.} The average of counterpoise corrected and uncorrected results was used for all contributions. In the case of the last contribution, the random walk procedure was kickstarted using the difference between the extrapolated result and the corresponding data in the haug-cc-pVTZ basis set. Moreover, we found that the Hartree-Fock contribution is practically converged within the basis set used and the uncertainty of this component is practically negligible. Therefore, only the remaining two components were taken into account in the uncertainty propagation scheme. The raw results for each component of this composite scheme are given in the Supplementary Material \cite{supp}.

\subsection{Results and discussion}

For each system in the S66 dataset, we calculated the interaction energies using the composite scheme defined in Sec.~\ref{subsec:composite_s66}. Next, the confidence intervals for each results were determined using the procedure proposed in this work. We consider three confidence levels, 75\%, 95\%, and 99\%, in these simulations. The confidence level of 75\% can be seen as rather loose, but it can be useful, e.g. at an exploratory stage of some applications. The 95\% confidence level is likely adequate for most practical uses since experimental errors in many fields are also reported at a comparable level. Indeed, for uncertainties originating from the normal (Gaussian) distribution, such as random error in repeated measurements, it corresponds very closely to the $2\sigma$ error bars. Finally, the 99\% confidence level may be useful especially in high-accuracy calculations where the reliability of the data is of prime importance.

In Tables~\ref{tab:s66_1}~and~\ref{tab:s66_2} we report the results of these calculations (the separation into two tables was done for editorial reasons and carries no other meaning). To simplify the discussion, we call the uncertainty prediction successful for a given system (at a particular confidence level) if the determined error bars for the result of the composite scheme cover the reference value. Otherwise, i.e. if the reference point lies outside the determined error bars, we say the uncertainty prediction is unsuccessful.

From Tables~\ref{tab:s66_1}~and~\ref{tab:s66_2} we see that the uncertainty prediction at the 75\% certainty level is successful in 46 out of 66 cases, a success rate of roughly 70\%. Moving to the 95\% confidence level, we have 63 successful predictions (approximately 95\% actual success rate). Finally, at the 99\% confidence level we have two unsuccessful predictions, giving the success rate of 97\%. However, it is worth pointing out that in these two unsuccessful cases (system 53 and 66) the reference value lies by only about 0.002 kcal/mol and 0.005 kcal/mol, respectively, from the edge of the 99\% confidence interval. Therefore, we cannot exclude that the error of the reference data becomes substantial at this level which slightly skews the statistics. Note that in an ideal situation, we would expect the uncertainty prediction, e.g. at the 75\% confidence level to be successful exactly 75\% of the time. However, this requires the sample size to be infinite, and deviations from this idealized scenario are expected for a finite number of examples. Nonetheless, these idealized ratios are fairly close to the actual success rates we observe for the S66 dataset.

Based on the discussed success rates, we see clearly that the method proposed in this work is more reliable and also provides reasonably tight confidence intervals. Additionally, our method naturally leads to, in general, asymmetric error bars that take into account error cancellations between components (whenever they are statistically likely).

\section{Conclusions}

In this work, we considered the problem of estimating the uncertainty of a theoretical result obtained by the means of a quantum-chemistry composite scheme. We assume that each component of the composite scheme carries its own uncertainty, originating primarily from the basis set incompleteness error, and it can be found using methods available in the literature. The problem is to find the uncertainty attached to the sum of all components.

We begin by considering the simplest and most frequently used approach to this problem: summing the squares of the uncertainties of each component and taking the square root. Under the hood, this method assumes that each component of the composite scheme can be treated as an independent variable in the statistical sense. By considering an example of a composite scheme available in the literature, we show that this assumption is not well justified in practice.

Next, we formulate a new method, building upon the random walk approach to estimating the basis set incompleteness error, to attach an uncertainty of a composite result without any additional assumptions. The proposed method is based on a series of random walks, initiated from randomized different starting values that represent the possible variation of the CBS limits of the available component data. We first illustrate the outcome of the method on two relatively simple examples of water dimer and nitrogen molecule. This enables us to outline and understand the most salient features of the method. Next, we apply the proposed uncertainty estimation scheme to a wider set of systems from the S66 database for which we design an example of a composite scheme. We show that the uncertainty estimates are reliable, yet tight, and can be useful in practice. Critically, the uncertainties carry a clear statistical meaning, corresponding to a predefined confidence level.

We believe that the proposed method can be used more broadly in computational physics and chemistry in various situations where several sources of error in the calculated data have to be taken into account simultaneously and the statistical relation between them is not known \emph{a priori}.

\section*{Supplementary Material}

See the Supplementary Material for the structure of the water dimer used in Sec.~\ref{sec:uqcomposite} and for raw interaction energies used in the composite scheme from Sec.~\ref{subsec:composite_s66}.

\begin{acknowledgments}
We are grateful to Prof. Rafał Latała (University of Warsaw) for fruitful discussions about the properties of the distribution $\rho_Y$. This work was supported by the National Science Centre, Poland, under research project 2024/54/E/ST4/00253. We gratefully acknowledge Poland’s high-performance Infrastructure PLGrid (HPC Centers: ACK Cyfronet AGH, PCSS, CI TASK, WCSS) for providing computer facilities and support within computational grants PLG/2025/018692.
\end{acknowledgments}

\section*{Data Availability Statement}

The data that support the findings of this study are available within the article and its supplementary material. A~Python implementation of the uncertainty prediction method proposed in this work is openly available: \url{https://github.com/lesiukmichal/uncertainties-composite}. This repository additionally includes sample input files and basic usage instructions.

\appendix

\section{\label{sec:appendix} Properties of the universal distribution $\rho_Y$}

The objective of this appendix is to provide a detailed derivation of several properties of the probability distribution $\rho_Y$. For convenience, we assume that the starting values are $y_0=0$ and $y_{-1}=-1$. This shift of the lower index in comparison with the general form given in the main text does not affect the limiting value $y_\infty$ we are interested in. Each term $y_{n+1}$ in the random walk procedure is generated according to the following expression:
\begin{align}
    y_{n+1} = y_n + u_n|y_n-y_{n-1}|, \;\;\; n\geq 0,
\end{align}
where $u_n$ is a random variable independent of $Y$, drawn from the uniform distribution $U$ in the interval $[-1,1]$. The above expression is just a restatement of the bound given in Eq.~(\ref{eq:fundamental-bound}). One can show by induction that the following expression then holds:
\begin{align}
    y_{n+1} = u_0 + u_0\,u_1 + u_0\,u_1\,u_2 + \ldots + u_0\,u_1 \cdots u_n,
\end{align}
where the starting term $y_0$ does not appear because $y_0=0$. In the above expression, each individual $u_i$, $i=1,2,\ldots,n$, is again a random variable with the uniform distribution $U[-1,1]$. Crucially, $u_i$ are independent of each other, because each step in the random walk is based on a fresh draw from the uniform distribution. Let us now take the limit $n\rightarrow\infty$ in the above expression. This leads to a formally infinite sum:
\begin{align}
    y_\infty = u_0 + u_0\,u_1 + u_0\,u_1\,u_2 + \ldots + u_0\,u_1 \cdots u_n + \ldots
\end{align}
Next, we rewrite the above expression in terms of probability distributions rather than the random variables. The probability distribution of $y_\infty$ is, by definition, $Y$, while the random variable representing each step is $U_i$. This leads to the equivalent formula:
\begin{align}
\label{eq:seriesy}
    Y = U_0 + U_0\,U_1 + U_0\,U_1\,U_2 + U_0\,U_1\,U_2\,U_3 + \ldots
\end{align}
Let us now take the factor $U_0$ out of the sum:
\begin{align}
    Y = U_0( 1 + U_1 + U_1\,U_2 + U_1\,U_2\,U_3 + \ldots ).
\end{align}
As the summation in the above expression is infinite and each distribution $U_i$ is identical, the distribution in the parentheses is just $1+Y$. This leads to the following equation for the probability distribution $Y$:
\begin{align}
\label{eq:master1}
    Y = U(1+Y),
\end{align}
where we have dropped the subscript for clarity. Note that this formula exactly coincides with Eq.~(\ref{eq:master}) from the main text. Additionally, $U$ and $Y$ are independent random variables here. This can be proven, e.g. by showing that the covariance of $U$ and $Y$ is zero by inserting the series from Eq.~(\ref{eq:seriesy}) into the definition of $\mbox{Cov}(U,Y)$.

On the surface level, Eq.~(\ref{eq:master1}) looks simple, but its actual content is actually rather complicated. In principle, one can use the conventional expressions for the sum and product of two probability distributions, and turn the above formula into an explicit integral equation for $\rho_Y$. However, despite considerable effort, we have not managed to solve this equation, and the result does not seem to be expressible through the standard elementary and/or special functions. Therefore, to learn more about $\rho_Y$ we take a different route based on the characteristic function which we describe next.

Let us recall that the characteristic function $\phi_Y$ of a probability distribution $\rho_Y$ is defined as the expectation value:
\begin{align}
    \phi_Y(t) = \mathbb{E}\big( e^{itY} \big).
\end{align}
or equivalently as the inverse Fourier transform:
\begin{align}
\label{eq:phifourier}
    \phi_Y(t) = \int_{-\infty}^{+\infty} dy\;\rho_Y(y)\,e^{ity}.
\end{align}
Note that $\phi_Y(0)=1$ which is a universal property of the characteristic function of any normalized distribution. Taking the expectation value $\mathbb{E}$ of both sides in Eq.~(\ref{eq:master}) leads to:
\begin{align}
    \phi_Y(t) = \mathbb{E}\big( e^{itU(1+Y)} \big).
\end{align}
We will evaluate the expectation value on the right in two steps, taking advantage of the fact that $U$ and $Y$ are independent. First, we will evaluate the expectation value over the domain of $Y$ treating $U$ as a constant (conditional probability), and then evaluate the expectation value over the domain of $U$. This leads to:
\begin{align}
\begin{split}
    &\mathbb{E}\big( e^{itU(1+Y)} \big) = 
    \mathbb{E}_U \Big( \mathbb{E}_Y\big( e^{itU(1+Y)} \big) \Big) = \\
    &\mathbb{E}_U \Big( e^{itU}\mathbb{E}_Y\big( e^{itUY} \big) \Big) =
    \mathbb{E}_U \Big( e^{itU}\phi_Y(tU) \Big),
\end{split}
\end{align}
where the subscripts have been added to the symbols of the expectation values to make clear over which variables they are evaluated. Rewriting the above expression explicitly in the integral form gives:
\begin{align}
    \phi_Y(t) = \frac{1}{2}\int_{-1}^{+1} du\;e^{itu}\phi_Y(tu).
\end{align}
As the distribution $\rho_Y$ is symmetric around zero, the same must be true for $\phi_Y$. Exploiting this parity, one can simplify the above integral to the form:
\begin{align}
    \phi_Y(t) = \int_{0}^{1} du\;\cos(tu)\phi_Y(tu).
\end{align}
After a simple exchange of variables to $x=tu$ we finally obtain the following integral equation for the characteristic function:
\begin{align}
    \phi_Y(t) = \frac{1}{t}\int_{0}^{t} dx\;\cos(x)\phi_Y(x).
\end{align}
Let us now multiply both sides by $t$ and take the derivative with respect to the variable $t$. This leads to the differential equation:
\begin{align}
    t\,\phi'_Y(t) + \phi_Y(t) = \cos(t)\phi_Y(t),
\end{align}
and after simple rearrangements one obtains:
\begin{align}
    \frac{\phi_Y'(t)}{\phi_Y(t)} = \frac{\cos t - 1}{t}.
\end{align}
Integrating both sides of this formula within the interval $t\in[0,x]$ gives:
\begin{align}
    \ln \phi_Y(x) - \ln \phi_Y(0) = -\int_0^x dt\;\frac{1 - \cos t}{t}.
\end{align}
From the above discussion we know that $\phi_Y(0)=1$, so the second term on the left-hand side vanishes. Moreover, the integral on the right-hand side is a known special function: the cosine integral, $\mbox{Cin}(x)$. After taking the exponential of both sides we finally obtain:
\begin{align}
    \phi_Y(x) = e^{-\mbox{\scriptsize Cin}(x)}.
\end{align}
The probability distribution $\rho_Y$ can be now recovered by recalling Eq.~(\ref{eq:phifourier}) and taking the Fourier transform:
\begin{align}
\begin{split}
    &\rho_Y(y) = \frac{1}{2\pi} \int_{-\infty}^{+\infty} dt\;\phi_Y(t)\,e^{-ity} = \\
    &\frac{1}{\pi} \int_{0}^{+\infty} dt\;e^{-\mbox{\scriptsize Cin}(t)}\,\cos(ty),
\end{split}
\end{align}
where in the last step we again used the parity of the characteristic function. The resulting equation coincides with Eq.~(\ref{eq:rhoint}) from the main text. 

This integral representation can be used as a starting point for deriving numerous other properties of the distribution $\rho_Y(y)$. For example, by taking the derivative with respect to $y$ and then integrating the resulting expression by parts, it is possible to find the following differential equation:
\begin{align}
\label{eq:rhoydiff}
    2y\rho_Y'(y) + \rho_Y(y+1) + \rho_Y(y-1) = 0.
\end{align}
It is reminiscent of, but distinct from, the Dickman differential equation \cite{dickman30}, which defines the Dickman function used in the prime number theory. Unfortunately, the differential equation~(\ref{eq:rhoydiff}) is simultaneously advanced and delayed due to the presence of $y-1$ and $y+1$ terms. This makes the application of the standard grid-based numerical solution methods extremely numerically unstable. Because of that, we found that the integral representation in Eq.~(\ref{eq:rhoint}) is a much more convenient starting point for numerical evaluation. However, the differential equation~(\ref{eq:rhoydiff}) is useful for deriving a small $y$ expansion of $\rho_Y(y)$ that was mentioned in the main text. By inserting Eq.~(\ref{eq:rhoint}) into Eq.~(\ref{eq:rhoydiff}) we found the following values of the coefficients:
\begin{align*}
\begin{split}
    &a_0 = -\rho_Y(1) \approx -0.178718, \\
    &b_0 = \lim_{y \to 0} \left[ \rho_Y(y) + \rho_Y(1)\ln|y| \right] \approx 0.275880, \\
    &a_2 = \frac{1}{8} \rho_Y(1) \approx 0.0223398, \\
    &b_2 = \frac{1}{8} \left[ \rho_Y'(2) - \rho_Y(2) - \rho_Y(1) - b_0 \right] \approx -0.0531882.
\end{split}
\end{align*}
To evaluate all of these quantities (aside from $b_0$) one requires only values of the distribution $\rho_Y(y)$ (and its first derivative) at points away from $y=0$. Therefore, they can be evaluated purely numerically as described in the main text, employing only the integral representation Eq.~(\ref{eq:rhoint}) and its derivative, where the quadrature is sufficiently numerically stable. The only exception is the coefficient $b_0$. However, in this case the $y\rightarrow0$ limit in the above expression can be worked out analytically, leading to:
\begin{align}
\begin{split}
    &b_0 = \frac{1}{\pi} \int_1^\infty dt\,\left[ e^{-\text{Cin}(t)} - \frac{e^{-\gamma}}{t} \right] \\
    &+ \frac{1}{\pi} \int_0^1 dt\;e^{-\text{Cin}(t)} - \frac{\gamma e^{-\gamma}}{\pi},
\end{split}
\end{align}
where $\gamma$ is the Euler-Mascheroni constant. The above integrals are evaluated numerically using exactly the same approach as in the case of Eq.~(\ref{eq:rhoint}).

As a final remark, we note that some other properties of the distribution $\rho_Y(y)$ can be found more conveniently starting from Eq.~(\ref{eq:master1}) rather than using the integral representation of Eq.~(\ref{eq:rhoint}). For example, knowledge of the moments of this distribution, i.e. $\mathbb{E}\big( Y^n \big)$, $n=1,2,3,\ldots$, can be useful in some applications. Since $\rho_Y(y)$ is symmetric around $y=0$, only even-$n$ moments are non-zero. By squaring both sides of Eq.~(\ref{eq:master1}) we obtain $Y^2=U^2(1+Y)^2$ and the evaluation of the expectation value of this formula leads to:
\begin{align}
\begin{split}
    &\mathbb{E}\big( Y^2 \big) = \mathbb{E}\big( U^2(1+Y)^2 \big) = 
    \mathbb{E}\big( U^2\big)\cdot \mathbb{E}\big( (1+Y)^2 \big) = \\
    &\mathbb{E}\big( U^2\big) \left[ 1 + \mathbb{E}\big( Y^2 \big) \right],
\end{split}
\end{align}
where we have used the fact that $U$ and $Y$ are independent variables, as well as the identities: $\mathbb{E}\big( 1\big)=1$ and $\mathbb{E}\big( Y\big)=0$. One can check that the second moment of the uniform distribution $U$ is $\mathbb{E}\big( U^2\big)=\frac{1}{3}$. This leads to the conclusion that $\mathbb{E}\big( Y^2\big)=\frac{1}{2}$. A similar procedure can be applied to derive higher-order moments and it leads to the following general recursion relation:
\begin{align}
    \mathbb{E}\big(Y^{2n}\big) = \frac{1}{2n} \sum_{i=0}^{n-1} \binom{2n}{2i} 
    \mathbb{E}\big(Y^{2i}\big).
\end{align}
However, we have not managed to find an explicit closed-form expression for $\mathbb{E}\big( Y^n \big)$ for even $n$.

\bibliography{uq}

\end{document}